\documentclass[twocolumn,prb,citeautoscript,superscriptaddress]{revtex4-2}
\UseRawInputEncoding
\usepackage{graphicx}
\usepackage{gensymb}
\usepackage{color}
\usepackage[dvipsnames]{xcolor}
\usepackage{float}
\usepackage{upgreek}
\usepackage{bm}
\usepackage{mhchem}
\usepackage{amsmath,amssymb}

\begin{document}

\title{Accurate noise spectroscopy using quantum sensing ensembles}
\author{Dhilan T. Vallury}
\affiliation{School of Physics, Faculty of Science, University of Melbourne, Melbourne, VIC, Australia}
\affiliation{Australian Research Council Centre for Quantum Biotechnology, School of Physics, University of Melbourne, Melbourne, VIC, Australia}

\author{Nikolai Dontschuk}
\affiliation{School of Physics, Faculty of Science, University of Melbourne, Melbourne, VIC, Australia}

\author{Alexander M. Jakob}
\affiliation{School of Physics, Faculty of Science, University of Melbourne, Melbourne, VIC, Australia}
\affiliation{Australian Research Council Centre for Quantum Computation and Communication Technology, School of Physics, University of Melbourne, Melbourne, VIC, Australia}

\author{Alexander J. Healey}
\email{alexander.healey2@rmit.edu.au}
\affiliation{Department of Physics, School of Science, RMIT University, Melbourne, VIC, Australia}

\author{David A. Simpson}
\email{simd@unimelb.edu.au}
\affiliation{School of Physics, Faculty of Science, University of Melbourne, Melbourne, VIC, Australia}
\affiliation{Australian Research Council Centre for Quantum Biotechnology, School of Physics, University of Melbourne, Melbourne, VIC, Australia}

\begin{abstract}
    Noise spectroscopy with quantum sensors is a powerful tool for characterising the properties of the sensor environment and detecting critical magnetic phenomena. While replacing single sensors with large ensembles promises substantial gains in measurement sensitivity and field of view, the output becomes difficult to interpret when there is a variable coupling to targets across the sensing ensemble. In this work we present an analysis procedure for accessing the true source noise spectrum using ensemble coherence from dynamical decoupling measurements. We show that sequence timing plays a dual role in ensemble noise spectroscopy, setting both the frequency selectivity and the weighted collection of signal across the ensemble, and demonstrate how incorporation of the coupling distribution allows accurate parameter extraction of the noise. We apply this procedure to a model system, shallow nitrogen-vacancy centres in diamond, and show that consistent information about the spectral character of MHz-range surface noise can be obtained from two sensing ensembles featuring distinct depth distributions. Our results pave the way for using large quantum sensor ensembles for quantitative, broadband noise spectroscopy, with straightforward extension to other solid-state and molecular sensing platforms.
\end{abstract}

\maketitle

\section{Introduction}
The decoherence of quantum systems is dictated by the environmental noise sources they are exposed to. Dynamical decoupling techniques, which operate by manipulating the quantum system so as to narrow its sensitivity to a small window of the overall noise power spectrum, have been successful in improving coherence properties across numerous platforms~\cite{cywinski2008,kotler2011,muhonen2014}. Inverting the role of the quantum system to act as a sensor, these same tools can be used to map the spectral characteristics of its environment in a technique referred to as noise spectroscopy ~\cite{alvarez2011,bylander2011,chan2018}. For single-qubit sensors, the relation between the measured decoherence and the underlying noise spectral density is well understood \cite{bylander2011, alvarez2011}, enabling quantitative reconstruction of the environment source spectrum. Here the filter window set by the decoupling sequence can be swept to measure the spectral character of the sensing target, holding promise for investigating criticality in condensed matter systems ~\cite{mclaughlin2022,rovny2024,ziffer2024}. For instance, recent theoretical work has shown that subtle changes in spectral profile can provide unambiguous evidence of two-dimensional XY models of magnetism~\cite{potts2025} and superconductivity~\cite{curtis2024}.

Measuring these increasingly subtle changes in weak signals demands greater sensitivity than is typically accessible using single-qubit probes, motivating the use of sensing ensembles instead~\cite{taylor2008,bar-gill2012,xue2026,ziffer2024}. 
However, spins within that ensemble will in general couple inhomogeneously to the target of interest, for instance due to a distribution of distances to the target. This distribution presents complications for the interpretation of decoherence curves, with ensemble sampling leading to stretched exponential decays that are not easily invertible~\cite{johnston2006,ziffer2024}. 
Furthermore, since decay rates will in general depend on both frequency and distance, there is the risk of conflating the two responses and even mistaking a change in how the ensemble is sampled for properties of the noise. These complications have to date precluded sensitivity gains from sensing ensembles being realised in systems where an unknown spectral response is of interest. Extending ensemble noise spectroscopy to these applications necessitates an analysis pipeline that fully disentangles the coupling distribution from the true spectral response and is compatible with the high-order dynamical decoupling sequences that bring the requisite sensitivity and spectral resolution. 

In this work we develop and experimentally demonstrate a general framework for accurate noise spectroscopy with inhomogeneous sensing ensembles. We show that sequence timing plays a dual role in these systems: it not only sets frequency selectivity but also imposes a sampling bias of the signal across the sensor distribution. When ignored, these effects blur the two responses and ultimately obscure the resolved noise spectrum. We show that uninformed interpretations of ensemble data result in significant errors, and introduce an analysis framework to recover meaningful information by inserting knowledge of the coupling distribution. Simulating the response of a model system, shallow nitrogen-vacancy (NV) ensembles in diamond, we benchmark this procedure against realistic experimental noise as well as different forms of coupling distributions and noise spectra. Finally, we apply the methodology in experiment, taking NV ensembles with distinct depth profiles and showing that common diamond surface noise signatures can be reconstructed.

\section{Results}

\begin{figure}[t!]
    \centering
    \includegraphics[width=\linewidth]{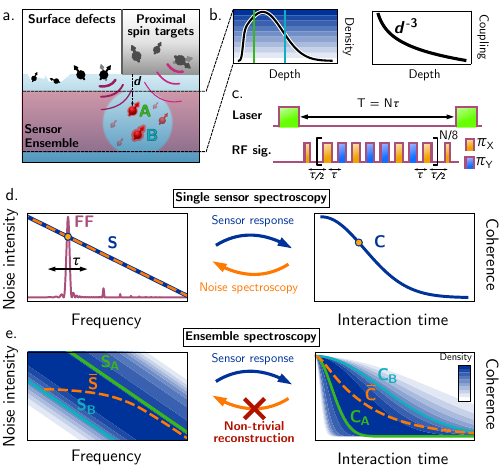}
    \caption{\textbf{Ensemble noise spectroscopy} \textbf{a.} Shallow sensor ensembles experience surface-limited coherence due to couplings to spin noise intrinsic to the surface (left), near to the interface (right), or both. \textbf{b.} The near-surface ensemble is characterised by a depth distribution (left) which sets sensor coherence via a $d^{-3}$ fall-off in coupling strength with depth (right).  \textbf{c.} XY8-N pulse protocol measuring NV coherence. A series of $\pi_{(X,Y)}$ pulse blocks totaling $N$ individual pulses refocus dephasing effects over a total interaction time $T=N\tau$. \textbf{d.} For a single sensor coupling a source noise spectrum $S$, performing XY8-$N$ with sweeping $\tau$ (and hence, $T$) provides the evolution of sensor coherence $C(T)$ for a given $N$. With knowledge of the sequence filter function (FF; purple), noise spectroscopy allows $S$ to be reconstructed (orange dashed) from measured coherence. \textbf{e.} Variable noise intensity across a sensor ensemble, arising from its depth distribution and depth-dependent coupling strength, precludes direct reconstruction of $S$ from ensemble coherence ($\bar{C}$) via the same noise spectroscopy approach. Instead an ensemble-modified spectrum $\bar{S}$ is determined, which presents deviant shape.}
    \label{fig1}
\end{figure}

The system we consider is an NV sensor some distance $d$ from a noise source at the diamond surface, illustrated in Fig.~\ref{fig1}a. That noise source can be intrinsic to the surface itself~\cite{romach2015,janitz2022} or the sensing target. We take the ensemble to have a depth distribution (Fig.~\ref{fig1}b), for instance produced by ion implantation statistics. Each NV will experience a noise spectrum taken as the separable product $S(\omega, d) = S'(\omega)D(d)$. $S'(\omega)$ is the spectrum characteristic of the source and $D(d)$ is the scaling of that signal with distance, which we take in all of the following to be $d^{-3}$ (See Supplementary Section III.D for further details). This spectrum can be accessed experimentally by measuring the decay in NV coherence $C(T)$ under a dynamical decoupling sequence with a known frequency response. In this work we consider sequences of the XY/CPMG form (Fig.~\ref{fig1}c), which have a known filter function set by the sequence timing \cite{bylander2011}; both the inter-pulse spacing $\tau$ and number of refocusing $\pi$-pulses $N$ set the total evolution time $T=N\tau$. The filter function decouples the sensor from noise off-resonant to the RF modulation rate $1/\tau$, defining a spectral window that may be scanned by varying $T$ (Fig.~\ref{fig1}d). Decoherence is set by the overlap between this window and the noise spectrum coupling to the sensor. For pure dephasing noise, a delta-like approximation of the selection window gives a straightforward inversion which maps the measured coherence of the sensor to the coupled noise \cite{szankowski2017}:
\begin{equation}
    S(\omega)=- \frac{\pi^2}{4T}\ln C(T).
    \label{eqn:S}
\end{equation}

For an ensemble sensor featuring some distribution $P(d)$ of NVs, varying noise spectra $S(\omega,d)$ (arising from a depth-dependent coupling strength, $D(d)$) coupling to the constituent sensors leads to variable coherence across the population (Fig.~\ref{fig1}e). The measured quantity is the weighted sum of NV coherence across the distribution $\bar{C}(T)$ given by (see Supplementary Section I.A for details)
\begin{equation}
    \bar{C}(T) = \int_0^\infty \exp\left[ -\frac{4T}{\pi^2}S'(\omega)D(d')\right]P(d')dd'.
    \label{eqn:Cbar}
\end{equation}
One can use Eqn.~\ref{eqn:S} to obtain a representative ensemble spectrum $\bar{S}(\omega)$, however the relationship between this quantity and the true distribution of noise experienced by the ensemble (Fig.~\ref{fig1}e) is unclear. Examination of Eqn.~\ref{eqn:Cbar} shows that the sampling of $P(d)$ is uneven in interaction time $T$; the exponential term introduces a sharp cut-off as $T$ becomes large such that only sensors for which $d>\left(\frac{4T}{\pi^2}S'(\omega)\right)^{1/3}$ contribute significantly. In this sense the interaction time plays a dual role in ensemble noise spectroscopy, defining both the frequency and coupling distribution sampling, risking conflation of the two in subsequent analysis.

\begin{figure}
    \centering
    \includegraphics[width=\linewidth]{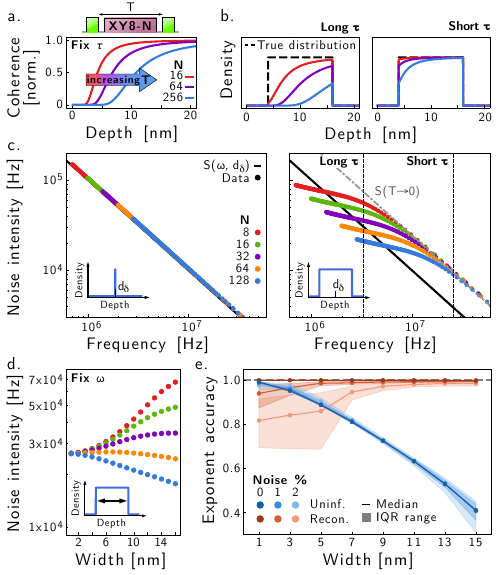}
    \caption{\textbf{Recovering accurate noise spectra}. \textbf{a.} Calculated sensor coherence across depth for a fixed inter-pulse spacing, $\tau$. Decoherence depends both on sensor depth and the total interaction time $T$, which scales with pulse number $N$. \textbf{b.} When considering the true sensor distribution explicitly, these dependencies lead to variable sub-ensemble sampling across $N$ when varying $\tau$. \textbf{c.} Measured noise spectra in the case of (left) a single NV (or equivalently $\delta$-layer) and (right) an NV ensemble evenly distributed across a finite depth range. \textbf{d.}  Coupled noise intensity at a fixed resonant frequency of $\omega = 4.0$~MHz for varying sequence length, N. Resolved values diverge with proportionality to distribution width \textbf{e.} Fitting accuracy comparison between an uninformed (blue) fitting of a source spectrum $S'$, and a `reconstruction' model (orange) which accounts for the underlying depth profile. The fitted power exponent of the known power-law shape defines the accuracy metric. Error margins represent statistical IQR from $N_\text{runs}=200$ (details in Supplementary Section II.B).}
    \label{fig2}
\end{figure}

This effect is evident in Fig.~\ref{fig2}a, where we consider XY8-$N$ dynamical decoupling sequences probing a fixed frequency $\omega=\pi/\tau$ and simulate the measured coherence across sensor depth for different $N$. As a consequence of the total interaction time $T$ scaling with $N$, coherent signal at longer sequence lengths is measured by a deeper sub-ensemble. Multiplying these functions by a test depth distribution for short and long $\tau$ (Fig.~\ref{fig2}b) gives an illustration of how ensemble sub-sampling can also vary with probe frequency.

To study the effect of this variable sub-sampling we take a $1/\omega$ noise spectrum that scales as $D(d)= d^{-3}$ and simulate the ensemble response to a series of XY8-$N$ sequences. The limiting case where there is no distribution of depths (i.e. for a single sensor or a true $\delta$-profile; Fig.~\ref{fig2}c, left) reproduces the expected spectra regardless of interaction time. When a finite rectangular sensor distribution is considered (Fig.~\ref{fig2}c right), two key features emerge. For long interaction times (low-$\omega$) the individual XY sequences are striated, leading to a distribution of points that is set more by the sensor distribution than the noise spectrum. For individual sequences, we also see a transition from the high-$\omega$ regime, where interaction time effects are negligible and all sequences faithfully reproduce the true $1/\omega$ of $S'(\omega)$ with an amplitude weighted by $P(d)$ (see Supplementary Section I.B for derivation of weighting), to a flattened gradient at low-$\omega$. This reflects the variable sampling of $P(d)$ as $\tau$ is swept, obscuring the true frequency dependence of $S'$. In this case the coupling distribution introduces an inflection in the data that could be erroneously identified as a cut-off frequency in the noise spectrum.

Fig. \ref{fig2}d shows explicitly how the spread in apparent noise spectral density scales with the width of the sensing ensemble. Here, a fixed source spectrum $S'$ couples to a rectangular sensor ensemble varying in width about a central depth $d_\delta=10$~nm, and the simulated $\bar{S}$ is evaluated at an arbitrary single frequency ($\omega=4.0$~MHz) for different $N$. In this case, distributions less than 3~nm wide are $\delta$-like while the striations become pronounced beyond this point. This increased spread at low-$\omega$ translates to an increasingly inaccurate fit of the noise spectrum when the depth distribution is not considered, even with precise knowledge of the spectral shape (Fig.~\ref{fig2}e). For instance, at a thickness of 7~nm the uninformed application of Eqn.~\ref{eqn:S} consistently underestimates $S'\propto \omega^{-0.8}$, a systematic error large enough to misidentify critical behaviour in magnetic systems~\cite{potts2025}. By considering the interaction time dependence explicitly in a fit of the simulated experimental data and assuming we know the sensor depth distribution perfectly (see Supplementary Section II.A for details), the true noise profile is able to be extracted with high confidence. This remains true even in the presence of readout noise up to 2\%.

\begin{figure}[ht!]
    \centering
    \includegraphics[width=\linewidth]{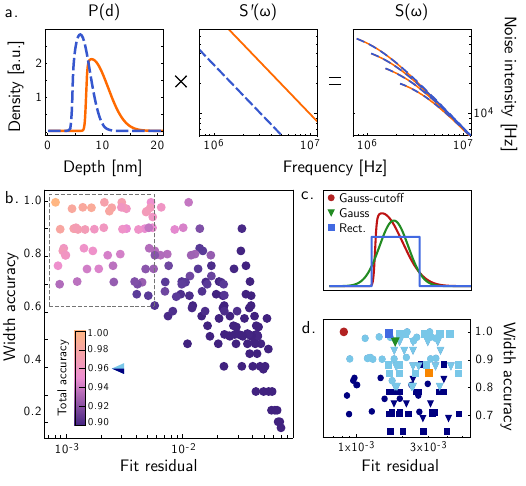}
    \caption{\textbf{Dependence of accuracy on distribution knowledge} \textbf{a.} Example of an ambiguity in ensemble measurement. \textbf{b.} Correlation between the `experimental' fit residual and width accuracy for guess depth distributions which vary in their accuracy to the truth (characterised by 90\% population about the median depth). Colours represent the multiplicative accuracy of the fitted power-law parameters (exponent and amplitude). \textbf{c.} Best-fit depth distributions for different profile shapes. \textbf{d.} Zoom of b. including all points exceeding 98\% exponent accuracy. Lighter points are those which also exceed 98\% accuracy in amplitude. Incorrect guesses for the depth distribution shape (Gaussian, rectangle) are also included under the same constraints, distinguished by marker shape. The best fit residual case for each shape is also distinguished by colour.}
    \label{fig3}
\end{figure}

We now consider more realistic cases, where the underlying NV depth distribution and form of the true noise spectrum may not be perfectly known. Starting again with the $1/\omega$ spectrum, we consider the impact of inserting an incorrect guess for the depth distribution on the accuracy of parameter extraction. As a distinguishability criterion we can consider the residual between the reconstructed and simulated experimental data. Fig.~\ref{fig3}a illustrates the ambiguity that arises when neither the depth distribution or noise spectrum are known: there is a set of indistinguishable solutions that trade off between the depth profile of the sensor ensemble and the overall magnitude of the noise spectrum (see Supplementary Section I.C). This ambiguity highlights the need to accurately characterise the sensor coupling distribution (we discuss methods for achieving this experimentally below).

Fig.~\ref{fig3}b explores the accuracy with which the spectral parameters defining a power-law source spectrum ($S'(\omega)=10^C\omega^{-\alpha}$) can be determined experimentally subject to imperfect characterisation of the coupling distribution. Here a range of Gauss-cutoff guess profiles reflecting realistic shallow NV ensembles~\cite{healey2021} are used to fit data simulated from a truth distribution (Fig.~\ref{fig3}a, orange). Informed by the sensitivity of ensemble spectra to width (Fig.~\ref{fig2}), we define an accuracy metric for comparing depth profiles based on distribution width (see Supplementary Section II.C for further details and alternative metrics). Each guess profile is used to fit a simulated truth dataset (with 0.1\% noise, see Supplementary Section II.D) for the parameters describing the source spectrum, $S'$. The clear anti-correlations between the fit residual and the accuracy metrics for the depth distribution guess ($y$-axis) and noise spectrum fit (colour map) indicate that a lower fit residual corresponds to more accurate $S'$ reconstruction. We also find however that it is possible to adequately fit experimental data with inaccurate depth distributions, reflecting the ambiguity highlighted in Fig.~\ref{fig3}a. To expand on this analysis we include guess cases using simpler variations (Gaussian and rectangle, Fig.~\ref{fig3}c) of the depth profile, displaying only those cases which exceed 98\% accuracy in the fitted power-law exponent (Fig.~\ref{fig3}d; see also Supplementary Section II.E). Points with lighter shading corresponding to guess cases which also exceed 98\% accuracy in the fitted spectral amplitude. The banding of these high-accuracy points highlights that accurate $S'$ reconstruction can be achieved by accurately constraining depth distribution width, even when the precise shape of the distribution is not known.

\begin{figure}[ht!]
    \centering
    \includegraphics[width=\linewidth]{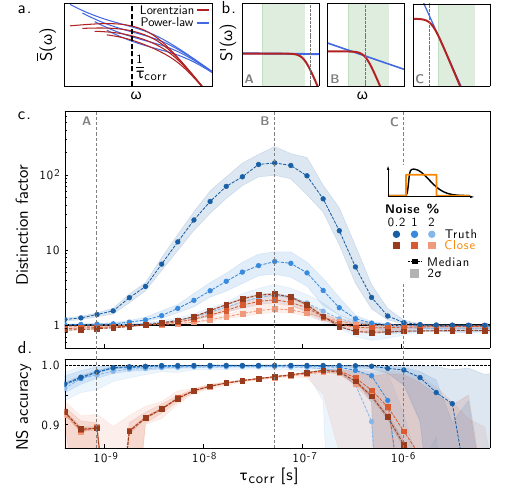}
    \caption{\textbf{Spectrum fitting with competing models} \textbf{a.} An example of best-fit $\bar{S}$ spectra when correctly (Lorentzian) and incorrectly (power-law) guessing the underlying $S$ form, in the case that the characteristic $1/\tau_\text{corr}$ inflection point sits within the measured range. \textbf{b.} Corresponding source spectrum $S'$ fits are shown in (B). Limiting cases where the inflection point instead occurs at higher (A) and lower (C) frequency to the measured range (green) are also shown. \textbf{c.} Simulation of the ability to distinguish true form of $S$ via $\bar{S}$ reconstruction using competing noise models. A distinction factor metric, defined as the ratio of incorrect (power-law) to correct (Lorentzian) $\chi^2$ residuals, is measured as the $1/\tau_\text{corr}$ value of the truth Lorentzian $S'$ is swept across the measured range. Two datasets are included: one using the known depth distribution (blue), and another using an approximation with incorrect form but good accuracy (orange; 85\% width accuracy). \textbf{d.} Corresponding total fit accuracy for each case.}
    \label{fig4}
\end{figure}

While the power-laws we have been considering so far do represent many physically relevant scenarios \cite{casola2018, chrostoski2021, potts2025}, it is also common to search for frequency cut-offs in e.g. Lorentzian spectra \cite{bar-gill2012,romach2015, hao2025}. We have seen that the true frequency scaling of the noise spectrum in experimental data can be heavily modified by a depth distribution, raising the question of whether we can identify true inflections in the noise spectrum. Fig.~\ref{fig4}a shows fits to data generated from a Lorentzian power spectrum, when the underlying shape (fit function) has been guessed correctly (red) and incorrectly (blue; power-law). Here the Lorentzian inflection point (often assigned to a bath correlation time, $\omega\sim1/\tau_{\text{corr}}$) is within the measured range and the two models are distinguishable through stratification and modified shape incurred by the depth distribution. Where the inflection point lies well outside of the measured range the two spectral forms are indistinguishable (Fig.~\ref{fig4}b).

\begin{figure*}[t!]
    \centering
    \includegraphics[width=\linewidth]{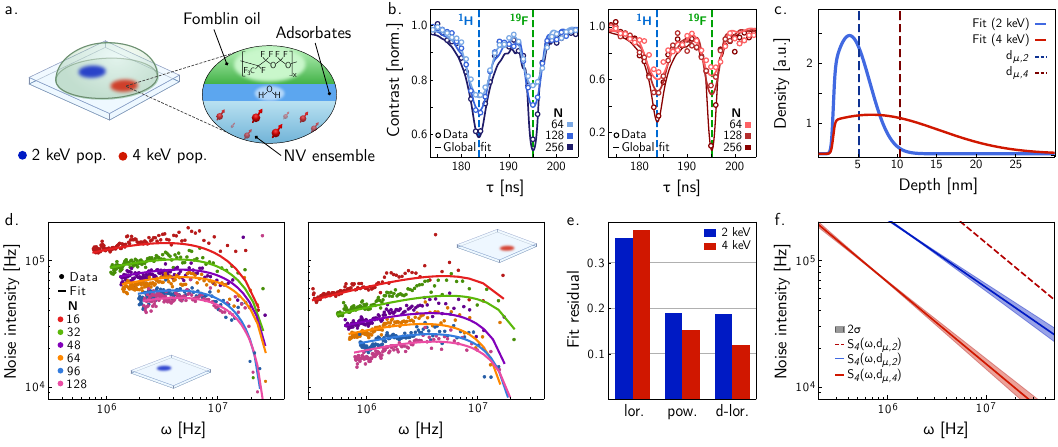}
    \caption{\textbf{Experimental demonstration of ensemble noise spectroscopy} \textbf{a.} Diamond substrate, host to two near-surface NV ensembles of varying implant energy (2 and 4~keV), is covered with Fomblin oil. \textbf{b.} NV-NMR of \ce{^1H} and \ce{^19F} on the controlled surface spin environment for the 2~keV (left) and 4~keV (right) NV ensembles. Data is fit across a series of XY8-$N$ curves for the global NV distribution. \textbf{c.} Fitted depth distributions (solid) for both ensembles. Mean depth, $d_\mu$, is also shown for both ensembles (dashed): $d_{\mu,2}$ = 5.09~nm and $d_{\mu,4}$ = 10.6~nm for the 2~keV and 4~keV distributions, respectively. \textbf{d.} Ensemble noise spectra $\bar{S}$ measured from the 2~keV (left) and 4~keV (right) ensembles are fit with their depth distributions constrained from NV-NMR measurements \textbf{e.} Fit residual comparison for common noise models. \textbf{f.} Reconstructed source spectrum $S$ from 2~keV ($S_2$, blue) and 4~keV ($S_4$, red) data fitting, evaluated at their respective mean depth $d_{\mu}$. The corresponding amplitude of $S_4$ evaluated at $d_{\mu,2}$ is also shown (dashed).}
    \label{fig5}
\end{figure*}

To assess whether the shape of the noise spectrum can be accurately determined in the intermediate cases where inflections occur on the periphery of measured frequencies, we again turn to the experimentally-accessible fit residual. We define a distinction factor $R_\text{pow}/R_\text{lor}$, where $R$ is the $\chi^2$-residual of the power-law and Lorentzian noise spectrum fits respectively (see Supplementary Section II.F for further details). In Fig.~\ref{fig4}c, data is simulated from a Lorentzian $S'$ with varying characteristic $\tau_\text{corr}$ values, such that the inflection point is swept across the subsequent probe range of a series of XY8-$N$ sequences, and the distinction factor determined. We find that the residual comparison accurately identifies the true form of the noise spectrum when the cut-off lies well within the measured band. This remains true in the presence of noise up to 2\% and with the insertion of an imperfect (orange point of Fig.~\ref{fig3}d) depth distribution, albeit over a reduced frequency range. Fig.~\ref{fig4}d confirms that the resulting accuracy of the noise spectrum reconstruction is high within this distinguishable band (see Supplementary Section II.G for further details).

Finally, we demonstrate the procedure by applying it to experimental data. We assess its efficacy by comparison of two NV sensing ensembles created via ion implantation into the same diamond substrate~\cite{healey2026}. The two ensembles are produced by 2 and 4~keV nitrogen ion implantation with a fluence of $10^{12}$~cm$^{-2}$. The spin coherence of these ensembles should be dominated by similar surface noise profiles, with the magnitude of the coupling differing in correspondence to the two depth distributions.

We have seen that accurate fitting of the noise spectrum is predicated on the depth distribution being known with some certainty. Here we determine the depth distribution using NV-based nuclear magnetic resonance (NMR) measurements~\cite{pham2016}, exploiting the same principle of sequence-dependent sub-ensemble sampling (Fig.~\ref{fig2}a) and the known Hydrogen and Fluorine NMR noise spectra~\cite{ziem2019,healey2021}. We coat the diamond surface in Fomblin oil ($\rho_{F}=41$~nm$^{-3}$) (Fig.~\ref{fig4}a), creating an effectively semi-infinite fluorine bath compared to the nanoscale sensing volumes ($\propto d_\text{NV}^3$) of shallow sensors \cite{staudacher2013}. This layer caps a finite adsorbate water layer ($\rho_H=67$~nm$^{-3}$) inherent to clean diamond surfaces exposed to atmosphere \cite{rugar2015}.
Fig.~\ref{fig4}b shows a subset of NMR sequences from each ensemble (left; 2~keV, right; 4~keV) used to fit the underlying depth distribution. Motivated by implant simulations \cite{posselt1992, ziegler2010} and the ionisation of NV$^-$ very close to the surface~\cite{broadway2018a,mccloskey2022}, we free-fit the 2~keV data with a distribution defined by a Gaussian with a shallow cut-off. The 4~keV data is also free-fit, with its cut-off constrained by the 2~keV fit to benefit convergence for the deeper distribution (see Supplementary Section III.A for further details).

With knowledge of the depth distribution constrained via NV-NMR, we fit ensemble noise spectra measured from both NV ensembles (Fig.~\ref{fig4}d). The data spans frequencies from 1 to 10~MHz and constitutes six different XY8-$N$ sequences (see Supplementary Section III.B for measurement details). Despite pronounced curvature that may suggest a frequency cut-off when interpreting the spectrum directly, we find that the data for both ensembles is well fit by a power-law. Indeed comparing residuals (Fig.~\ref{fig4}e), we find that the residual resulting from a Lorentzian fit is approximately a factor of two greater than for a power-law fit. This matches well with the scale of distinction factor predicted by Fig.~\ref{fig4}b for $\sim 2\%$ noise. Double-Lorentzian fits, despite including more free parameters, do not produce significantly more successful fits to the data (see Supplementary Section III.C for full comparison). The reconstructed power-law source spectra $S'$ are shown in Fig.~\ref{fig4}f, where the intensity corresponds to that at the mean depth $d_\mu$ of each ensemble. The fitted exponents, $\alpha_{2}=0.54(2)$ and $\alpha_4=0.65(2)$ for the 2 and 4~keV ensembles respectively, are similar, reflecting the detection of a common surface noise profile. The difference in intensity between ensemble curves reflects the increased coupling strength experienced by the 2~keV ensemble which resides closer to the surface. To facilitate comparison, we also plot $S'_{4}$ evaluated at the mean depth of the 2~keV ensemble, $d_{\mu,2}$. In this case the 4~keV fit implies more intense surface noise, with the difference possibly explained by a greater relative prevalence of bulk noise sources for the deeper implant (see Supplementary Section I.D).

The agreement between the extracted noise spectra showcases the accuracy of the noise spectrum deconvolution method when the underlying distribution of coupling strengths is well characterised. The recovery of a power-law noise spectrum deviates from single NV results which resolve Lorentzian spectra~\cite{romach2015}, and likely reflects the lateral distribution of noise sources across the diamond surface. As such, the combination of single NV and ensemble results highlights an example of the emergence of a $1/f$-type noise profile from a distribution of Lorentzian sources \cite{shehata2023}.

\section{Conclusion}
This work introduces a quantitative noise spectroscopy framework tailored to quantum sensing ensembles, where variations in sensor-source coupling are unavoidable. We show that significant coupling distributions fundamentally change the sensor's apparent spectral response, potentially leading to erroneous interpretations of data when timing effects are not considered. Provided the coupling distribution is known with high confidence (i.e. capturing distribution width within 10\%), our results suggest that accurate parameter extraction is possible even in the presence of realistic experimental noise up to 2\%. We demonstrate the framework experimentally using two shallow NV ensembles in diamond of varying depth profiles, identifying a consistent power-law surface noise spectrum.

Beyond shallow NV ensembles, the description developed here is generalisable to other sensing platforms featuring coupling distributions, including other solid state spin ensembles and molecular sensors unevenly dispersed in biological environments~\cite{feder2025}. By dealing with interaction time effects, our framework permits the use of high-order dynamical decoupling sequences that can offer high spectral resolution and signal specificity through pulse sequence engineering~\cite{romach2019,choi2020}. These benefits can help enable reliable high sensitivity noise spectroscopy in biological and condensed matter systems. Finally, an intriguing extension of the current work would be to explicitly consider cases where the sensor-target distance additionally imposes momentum selectivity~\cite{potts2025}, providing access to spatial structure factors using ensemble sensors~\cite{kazi2026}. 

\section{Acknowledgements}
The authors acknowledge useful discussions with David Broadway. This work was supported by the Australian Research Council (ARC) through grants DP220102518 and DP250100973, and by the ARC Centre of Excellence in Quantum Biotechnology through grant CE230100021. DS acknowledges support via ARC Mid-Career Industry Future Fellowship (IM240100073).  A.M. Jakob acknowledges funding of the ARC Centre of Excellence for Quantum Computation and Communication Technology (CE170100012) and the US Army Research Office (Contracts No. W911NF-17-1-0200 and W911NF-23-1-0113).

\section{Methods}

\subsection{Diamond sample}
The diamond sample in this work was prepared from an Element Six “electronic grade” diamond wafer, and consists of an array of 100~$\mu$m spots implanted with a range of implant energies and a range of dosages. Pre-implantation preparation involved acid boiling (6~mL \ce{H_2SO_4} 96\% and 0.5~g \ce{NaNO_3}, boiling), Piranha Etching (3~mL \ce{H_2SO_4} 96\% + 1~mL \ce{H_2O_2} 37\%, 100$^\circ$C), then 5~minutes ultrasonication in Hexane, Toluene, IPA, and DI water in order. The sample was implanted on an in house Colutron system \cite{jakob2024} with a 7$^\circ$ rotation from the [001] sample normal, 0$^\circ$ rotation from the [110]-oriented sample edge. Uncertainty in implant fluence is estimated to be well below 5\%. After implantation, the sample went through a final cleaning step of 5~minutes ultrasonication in Hexane, Toluene, IPA and DI water. The sample was annealed in a forming gas environment \cite{healey2026}. Annealing was done in a Carbolite tube furnace equipped with quartz tube, and differentially pumped RGA. The forming gas mixture was 5\% hydrogen in nitrogen. Trace oxygen levels were confirmed to be below 10~ppm for the entire anneal. After annealing, the sample was treated in a UV ozone cleaner for 30~minutes to remove the hydrogen termination.

\subsection{Spin measurements}
Spin measurements were performed on a custom-build widefield microscope. A 532~nm laser (Quantum Opus) is used for NV initialisation and readout, with an input power of 250~mW. PL from an approximately 50$\times$50~$\mu$m area over which the laser spot is roughly uniform (intensity $\sim$10~kW~cm$^{-2}$) was collected onto an sCMOS camera (Andor Neo). RF signals for NV spin state control were delivered using a Rohde \& Schwarz SMBV100A signal generator IQ modulated by a Keysight P9336A arbitrary waveform generator with 1~ns time resolution. Delivery is facilitated by a gold omega-shaped resonator patterned on a glass coverslip onto which the diamond substrate is mounted. A PulseBlaster ESR-pro 500 MHz card controls the timing of RF pulse sequences, laser pulses, and camera triggers with a time resolution of 2~ns. Spin measurements were conducted under a bias field of approximately 70~mT aligned with one set of NV axes. For XY8 measurements, a reference is taken by inverting the phase of the final readout $\pi/2$ pulse relative to the signal sequence to cancel common-mode noise. All data shown is the normalised difference between camera exposures which differ by the readout pulse phase.

\bibliography{library}

\end{document}


\title{Supplementary Information for `Accurate noise spectroscopy using quantum sensing ensembles'}

\author{Dhilan T. Vallury}
\affiliation{School of Physics, Faculty of Science, University of Melbourne, Melbourne, VIC, Australia}

\author{Nikolai Dontschuk}
\affiliation{School of Physics, Faculty of Science, University of Melbourne, Melbourne, VIC, Australia}

\author{Alexander M. Jakob}
\affiliation{School of Physics, Faculty of Science, University of Melbourne, Melbourne, VIC, Australia}
\affiliation{Australian Research Council Centre for Quantum Computation and Communication Technology, School of Physics, University of Melbourne, Melbourne, VIC, Australia}

\author{Alexander J. Healey}
\email{alexander.healey2@rmit.edu.au}
\affiliation{Department of Physics, School of Science, RMIT University, Melbourne, VIC, Australia}

\author{David A. Simpson}
\email{simd@unimelb.edu.au}
\affiliation{School of Physics, Faculty of Science, University of Melbourne, Melbourne, VIC, Australia}

\maketitle
\tableofcontents
\section{Mathematical description}
\subsection{Surface-limited ensemble noise spectroscopy}
\noindent The coherence of a single sensor in a classical dephasing noise environment is given by \cite{Szankowski2017}
\begin{equation}
    C(T)=e^{-\chi(T)},
    \label{eq:singleC}
\end{equation}
where the decoherence functional $\chi(T)$,
\begin{equation}
    \chi(T)=\frac{1}{2}\int_{-\infty}^\infty \frac{d\omega}{2\pi}|\Tilde{f}_T(\omega)|^2S(\omega).
    \label{eq:dec_functional}
\end{equation}
is an overlap integral between the noise spectral density $S(\omega)$ coupling the sensor and the squared amplitudes of $\Tilde{f}_T(\omega)=\int dt\,e^{-i\omega t}f_T(t)$. The latter is the Fourier transform of the time-domain filter function $f(T)$ describing the phase switching of the sensor and the square of its amplitudes, $|\Tilde{f}_T(\omega)|^2$, is commonly referred to as the frequency-domain filter function. For the CPMG/XY8-$N$ sequences relevant to this work, defined by $\delta$-like phase switching at times $T_k=(k-\frac{1}{2})T/N$, the corresponding filter function is
\begin{equation}
    |\Tilde{f}_T^\text{CPMG}|^2=\frac{16}{\omega^2}\frac{\sin^4\frac{\omega T}{4N}}{\cos^2\frac{\omega T}{2N}}\sin^2\frac{\omega T}{2}.
\end{equation}
Equation \ref{eq:singleC} described how the coherence $C(T)$ of a quantum system after some interaction time, $T$, is determined by the coupled noise environment $S(\omega)$. In the surface-limited regime, where dephasing is driven dominantly by noise at the interface, we consider a separable form $S(\omega,d)=S'(\omega)D(d)$. Here, $S'(\omega)$ is the noise spectrum at the source and $D(d)$ is the variation in the intensity of noise with distance $d$ away from that source. We assume here that standoff does not influence the frequency response (shape) of the noise and that the intensity scales as $d^{-3}$ with depth (see Sec. \ref{sec:dscaling} for further discussion).

In the case where coherence is measured from an ensemble of surface-limited sensors, the ensemble coherence, $\bar{C}(T)$, is then the total coherent signal accumulated across the measured population:
\begin{equation}
    \bar{C}(T)=\int_0^\infty C(T,d')P(d')dd',
\end{equation}
where $C(T,d')$ is the coherence of a sensor at depth $d'$ and $P(d')$ is the probability density function describing the sensor distribution across depths (i.e. $\int_{0}^\infty P(d')dd'=1$). Evaluating the single sensor expression for $C(T,d')$ in terms of the noise (Eqn.~\ref{eq:dec_functional}) gives
\begin{equation}
    \bar{C}(T)=\int_0^\infty dd'P(d')\exp[-D(d')\chi'(T)],
    \label{eq:generalCBar}
\end{equation}
where $\chi'(T)$ represents the decoherence functional incurred by the source spectrum, $S'$, which under the $\delta$-approximation for the filter function is related as $\chi'(T)=4TS'(\pi N/T)/\pi^2$ (Eqn.~1, main text). The embedded $D(d)$ scaling encodes the variation in decoherence rate as the coupling distance is increased.

\subsection{Short-$T$ limit}
\noindent Beginning with the simplified point inversion from coherence $C(T)$ to coupled noise $S(\omega)$ (Eqn. 1, main text), we find that in limit of short interaction times,
\begin{equation}
    S(T\rightarrow0)=-\frac{\pi^2}{4T}\ln\bar{C}(T\rightarrow0),
    \label{eqn:SshortT}
\end{equation}
is indeterminate. So, by Taylor expansion of $\bar{C}$ to first order in T:
\begin{align}
    \bar{C}(T)&\approx1-\chi'(T)\int_0^\infty dd'P(d')D(d')\\
    \Rightarrow \ln\bar{C}(T)&\approx-\frac{4T}{\pi^2}S'\biggl(\frac{\pi N}{T}\biggr)\int_0^\infty dd'P(d')D(d'),
\end{align}
where in the last line we use the first order Taylor series relation $\ln(1-x)\approx-x$. Inserting this into our initial expression for noise in the short interaction time limit (Eqn.~\ref{eqn:SshortT}) gives
\begin{equation}
    \bar{S}(T\rightarrow0)\approx S'\biggl(\frac{\pi N}{T}\biggr)\int_0^\infty dd'P(d')D(d').
\end{equation}
The inversion of ensemble coherence $\bar{C}$ in the short interaction time limit reconstructs the spectral shape of the source spectrum $S'$ with a coupling strength that is weighted by the expectation value of the depth scaling $D(d)$ under the sensor distribution density $P(d)$.

\subsection{The coupling ambiguity problem}
\label{sec:couplingambiguity}
\begin{figure}[H]
    \centering
    \includegraphics[width=0.8\linewidth]{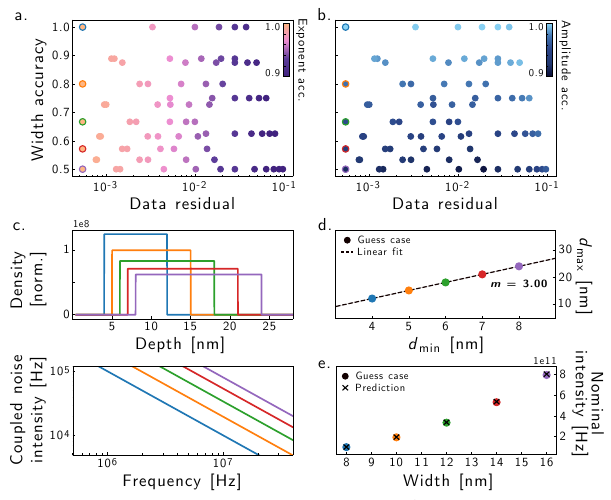}
    \caption{\textbf{a.} Exponent and \textbf{b.} Amplitude accuracy of fitting an ensemble $\bar{S}$ generated from a power law source spectrum ($S'$) measured using a rectangular distribution of sensors. Points of highest exponent accuracy ($>$99.5\%) are outlined. \textbf{c.} (top) Guess distributions for the high exponent accuracy cases, distinguished by colour. (bottom) Corresponding fitted power law source spectra. \textbf{d.} Scaling of the parameters defining the DD in high accuracy cases is well fit by a linear trend. \textbf{e.} Fitted amplitude of the nominal (source) spectrum compared to the width of the guess distribution. Simulation values (dots) are corroborated with the predicted value (crosses) determined by the scaling of $D(d)$.}
    \label{fig:couplingambiguity}
\end{figure}
\noindent To identify a degeneracy between the underlying depth distribution and the noise spectrum, let us consider the exponent within the integrand of Eq.~\ref{eq:generalCBar}:
\begin{equation}
    \bar{\chi}(T)= -\frac{4T}{\pi^2} D(d)\times S'\biggl(\frac{\pi N}{T}\biggr),
\end{equation}
where $D(d)\propto d^{-3}$ is determined by the depth distribution and $S'({\pi N}/{T})$ the source noise spectrum. We see that in general, a rescaling $S'\rightarrow\lambda S'$ can be directly compensated with the inverse scaling $D\rightarrow D/\lambda$. This renders the decoherence functional $\bar{\chi}$ invariant: simultaneously scaling the noise intensity $S'\rightarrow\lambda S'$ and depth profile $d'\rightarrow \lambda^{1/3}d'$ (i.e. $d_\text{min/max}\rightarrow \lambda^{1/3}d_\text{min/max}$) returns the same decoherence functional $\bar{\chi}$ as the unscaled system. Systems which map to one other in this manner are therefore degenerate, implying that the amplitude of $S'$ and the true scale (i.e. width) of $P$ cannot be disentangled from ensemble noise spectra without \textit{a priori} knowledge of either.

We demonstrate this problem clearly through a simulation using simple forms for the depth distribution and source noise spectrum. A truth dataset is generated using a rectangular sensor profile -- uniform density from $d_\text{min}=4$~nm to $d_\text{max}=12$~nm (width = $8$~nm) -- and a power law source spectrum $S'(\omega)=10^C\omega^{-\alpha}$ where in this case, $S(\omega,d)$ has parameter values $\alpha=1$ and $C=11$ at a nominal depth of $d_\text{nom}=8$~nm. Scatter plots measuring fitted accuracies are generated by the same procedure as described in Sec.~\ref{sec:scatterplots}, here for correct guesses for depth profile and noise spectrum shapes. In Fig.~\ref{fig:couplingambiguity}a, the cases of highest accuracy ($>$99.5\%) in the fitted exponent ($\alpha$) of the source spectrum are outlined and distinguished by colour. While the exponent accuracy of these optima are indistinguishably close to the true value, Fig.~\ref{fig:couplingambiguity}b reveals how these points are distinct in the fitted amplitude ($C$) accuracy. Nevertheless, this set a \textit{false optima} - wherein only one case (blue outline) represents the true system - fit the truth dataset the same (i.e. based on $\chi^2$ residual) and therefore cannot be distinguished experimentally.

Depictions for this set of guess depth distributions and their fitted source spectra are shown in Fig.~\ref{fig:couplingambiguity}c, depicting the interplay between the scaled depths of the sensor distribution (top) and the amplitude of the source spectrum (bottom). In Fig.~\ref{fig:couplingambiguity}d, we plot the two parameters ($d_\text{min}$, $d_\text{max}$) defining the simple rectangular profile for each optimum. We find that these points are fit well by a linear trend -- that is to say their ratio $d_\text{max}/d_\text{min}$, which is proportional to the width of the distribution, is constant. This is equivalent to a linear scaling $\lambda$ of depths from one case to another, which we have recognised are indistinguishable with a corresponding scaling $\lambda^3$ in the noise intensity of the source spectrum. To verify this, in Fig.~\ref{fig:couplingambiguity}e we compare the fitted amplitude value (dots) against the predicted value (crosses) which can be derived from the truth spectrum $S'_\text{truth}$ and relative scaling of distribution widths between cases. We find these two values corroborate almost exactly.

\subsection{Bulk contribution to environmental noise}
\label{sec:bulkcontribution}
Consider now the case where the noise environment consists of a `bulk-like' component. In other words, the source noise spectrum $S(\omega,d)$ can be decomposed as 
\begin{equation}
S(\omega,d)=S_\text{surf}(\omega,d)+S_\text{bulk}(\omega),
\end{equation}
where $S_\text{surf}$ is the depth-dependent surface noise contribution and $S_\text{bulk}$ is depth-independent contribution representing a homogeneously-coupled bulk component. The general form for ensemble decoherence (Eq. \ref{eq:generalCBar}) then becomes
\begin{equation}
    \bar{C}(T)=\int_0^\infty dd'P(d')\biggl[\exp\biggl\{ -\frac{1}{2}\int_{-\infty}^\infty \frac{d\omega}{2\pi}|\Tilde{f}_T(\omega)|^2\biggl(S_\text{surf}(\omega,d)+S_\text{bulk}(\omega)\biggr) \biggr\} \biggr]
\end{equation}
Since $S_\text{bulk}$ has no $d$-dependence, it factors as
\begin{equation}
    \bar{C}(T)=\exp\biggl(-\chi_\text{bulk}(T)\biggr)\int_0^\infty dd'P(d')\biggl[\exp\biggl\{ -\biggl(\frac{d_\text{nom}}{d}\biggr)^3\chi_\text{nom}(T)\biggr\} \biggr] \equiv C_\text{bulk}(T)\bar{C}_\text{surf}(T)
\end{equation}
The consideration of a bulk contribution manifests as a global exponential decay envelope that multiplies the original (surface-limited) $\bar{C}(T)$. Taking then the inversion to the coupled noise spectral intensity (Eqn.~1, main text), this bulk contribution manifests as a corresponding offset to the extracted $\bar{S}$ which is consistent across $N$:
\begin{align}
    \bar{S}(\omega)&=-\frac{\pi^2}{4T}\ln[C_\text{bulk}(T)\bar{C}_\text{surf}(T)]\\
    &=-\frac{\pi^2}{4T}\ln[C_\text{bulk}(T)] -\frac{\pi^2}{4T}\ln[\bar{C}_\text{surf}(T)]\\
    &= S_\text{bulk}(\omega) -\frac{\pi^2}{4T}\ln[\bar{C}_\text{surf}(T)].
\end{align}
If the resulting $\bar{S}$ curves were to be fitted with no bulk contribution and precise knowledge of the sensor distribution, the fitted source spectrum may be obscured in both its coupling intensity and spectral shape depending on the shape of the bulk component. This is a possible explanation for the discrepancy between resolved surface noise spectra between measured NV ensembles, where the deeper (4~keV) ensemble resolves a spectrum that is positively offset in amplitude (see Sec.~\ref{sec:allNSfits} for further discussion).

\section{Simulations}

\subsection{Minimisation fitting procedure for ensemble noise spectra}
The source noise spectrum $S'$ is determined experimentally by fitting Eqn.~2 (main text) to experimental data, using the Levenberg-Marquardt non-linear least-squares algorithm to obtain the parameters defining $S'$ for a chosen depth distribution $P$ and a function form for $S'$. The objective function for this procedure is defined by the residuals between input data (set on decoherence curves of varying $N$) $\{C_\text{N}^\text{input}\}$, and a guess dataset $\{C_\text{N}^\text{guess}\}$. The guess dataset is formed by evaluating ensemble coherence curves at $T_\text{int}$ values corresponding to the input dataset, using the function's input parameters. The objective function is parameterised by variables describing both $S'$ and $P$, however in practice the sensor distribution is constrained by other means (see Sec.~\ref{sec:NMR_DDfitting}) so that the fitted source spectrum is disambiguated (see Sec.~\ref{sec:couplingambiguity}). The residual is defined in the time-domain ($C_N^\text{input}-C_N^\text{guess}$) rather than in frequency-domain such that additive readout noise error, which scales harshly under the exponential inversion $S\propto \ln(C)/T$, does not strongly bias high-$f$ point residuals. The residuals are also normalised based on dataset length so that contributions from each $N$ in $\{C_N\}$ are equal.

\subsection{Complete statistics on reconstruction method}
\begin{figure}[H]
    \centering
    \includegraphics[width=\linewidth]{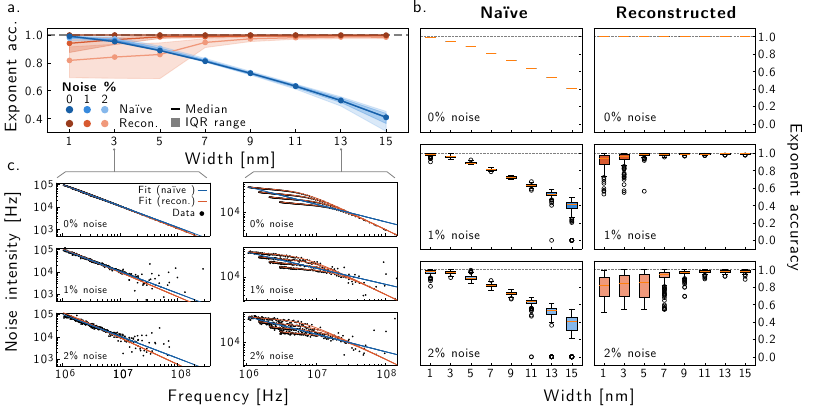}
    \caption{\textbf{a.} Accuracy comparison between naïve (blue) fitting of a source spectrum $S'$, and a `reconstruction' model (orange) which accounts for the underlying depth profile. The fitted power exponent of the known power-law shape defines the accuracy metric. Error margins represent the statistical interquartile range from $N_\text{runs}=200$. \textbf{b.} Complete box-plot statistics for each width case. Each case is performed $N_\text{runs}$ times with a randomised seed for noise to build statistics. \textbf{c.} Depiction of raw data and residual-minimised fits for a single run of noise in the small width (3~nm) and large width (13~nm) regimes.}
    \label{fig:fig2stats}
\end{figure}
Complete statistics on the accuracy of fitting methods for ensemble noise spectroscopy is shown in Fig~\ref{fig:fig2stats}. The fitted accuracy of the exponent defining a truth power-law source noise spectrum is compared between a direct interpretation  of the measured ensemble data, and a reconstruction method that considers the underlying sensor distribution. Shown in Fig~\ref{fig:fig2stats}a, accuracies are determined across a range of rectangular distribution widths for varying levels of noise. Values are determined from statistics on $N_\text{runs}=200$ randomly-seeded generations of noisy truth data, with the total statistics displayed as box-plots in Fig~\ref{fig:fig2stats}b. Explicit fits to ensemble data are shown for select cases in Fig~\ref{fig:fig2stats}c.

\subsection{Depth distribution accuracy metrics}
\begin{figure}[H]
    \centering
    \includegraphics[width=0.75\linewidth]{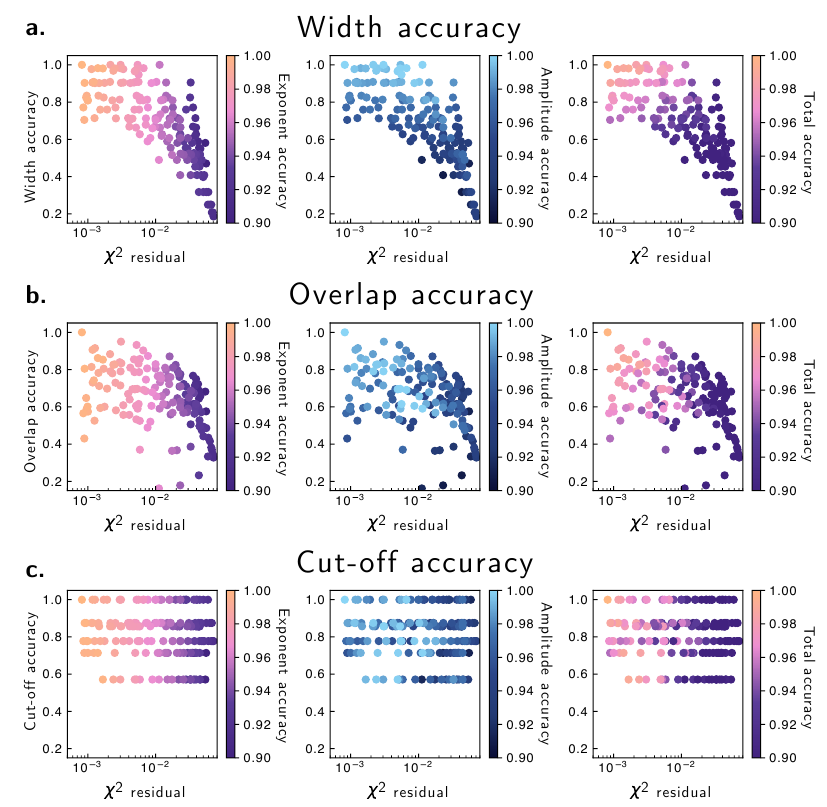}
    \caption{Scatter comparison between minimised fit residual and various accuracy metrics that describe the similarity of a guess depth distribution to the true profile. Metrics are defined based on the distribution's \textbf{a.} width, \textbf{b.} overlap, and \textbf{c.} cut-off value. Cases are plotted with colour gradient corresponding to the accuracy of fitted power-law parameters: left, exponent; middle, amplitude; right, multiplicative total.}
    \label{fig:ddmetrics}
\end{figure}
In the main text, we define an accuracy metric for comparing depth profiles based on distribution width. For generality in shape, the width of the distribution is defined as the depth range corresponding to 90\% of the total population, about the median value. The accuracy scatter plots under this metric is shown in Fig.~\ref{fig:ddmetrics}a. for each fitted power-law parameter, as well as their multiplicative total. Here we can resolve clear correlation (or lack thereof) between this measure and the fitted amplitude (or exponent) accuracy which is expected by the coupling ambiguity problem (see Sec. \ref{sec:couplingambiguity}). 

While Fig.~2 (main text) demonstrates that width is a defining characteristic of the sensor distribution for ensemble noise spectroscopy, it is not the only relevant metric. In Fig.~\ref{fig:ddmetrics}b, we analyse the same dataset under an alternative metric defined as the overlap between two normalised depth distributions. While the trend in amplitude accuracy is less clear, the multiplicative accuracy does generally improve with both higher overlap accuracy and smaller fit residual. This metric is distinct from width in that it is more susceptible to variation in shape: two distributions with very similar width may have less accurate overlap if they are peaked at different values, thereby reducing overlap under normalisation. Fig.~\ref{fig:ddmetrics}c shows a third metric based entirely on the cut-off value of the gauss-cutoff distributions. Here, the unclear optimal value and lack of trend is further indication that distribution width is an important characteristic defining the sensor profile.

\subsection{Scatter statistics for varying levels of noise}
\label{sec:scatternoise}
\begin{figure}[H]
    \centering
    \includegraphics[width=\linewidth]{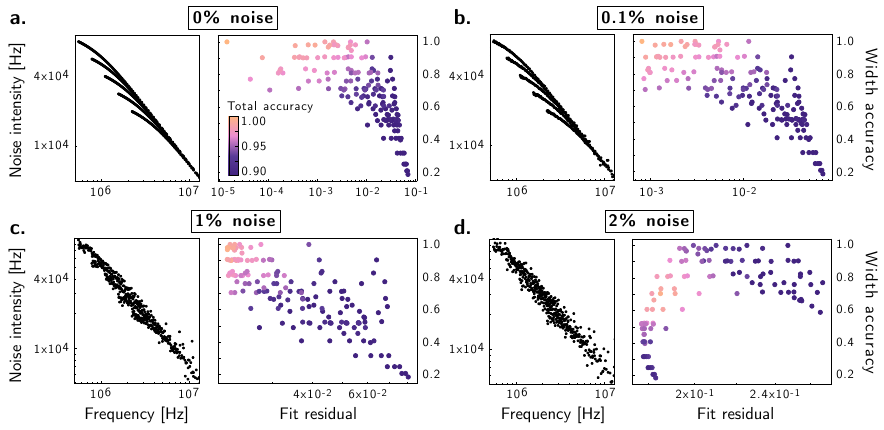}
    \caption{Fitting accuracies of ensemble noise spectra under different levels of simulated noise. In each noise case \textbf{a-d} the left panel shows the truth ensemble noise spectra, subject to noise, that is fit with a variety of guess sensor distributions to yield an accuracy scatter plot (right).}
    \label{fig:scatternoise}
\end{figure}
In Fig.~\ref{fig:scatternoise}, we show the effects that different levels of readout noise has on the particular case of the sensing ensemble system considered in Fig.~3 (main text). Coherence data is generated from a truth depth distribution $P$ and source spectrum $S'$, before stochastic readout noise (scaled by the noise percentage) is added to points along the curves. The corresponding ensemble spectra $\bar{S}$ form the truth dataset to be fit, and are shown on the left of each sub-figure (a-d) representing different levels of noise. This truth dataset is then fit for the parameters defining $S'$ (see Sec. II.A) using a variety of guess sensor distributions (see next section for further details). The resulting fitting accuracies are displayed as a scatter plot on the right of each sub-figure (a-d). In the noise-free case, exact guesses to the depth profile yield arbitrarily low residuals (Fig.~\ref{fig:scatternoise}a). Under 0.1\% noise (Fig.~\ref{fig:scatternoise}b), a comparable set of accurate (>95\%) guesses are distinguished by the fit residual from those with lower accuracy ($\sim$90\%) and the anti-correlation between the fit residual and width accuracy remains clearly resolved. Despite a significant rise in the noise floor at 1\% noise (Fig.~\ref{fig:scatternoise}c), this trend is still resolvable and only breaks down once the noise reaches a magnitude comparable to the striations between sequences (Fig.~\ref{fig:scatternoise}d), which is around 2\% noise in this case.

\subsection{Scatter statistics for different profile shapes}
\label{sec:scatterplots}
\begin{figure}[H]
    \centering
    \includegraphics[width=\linewidth]{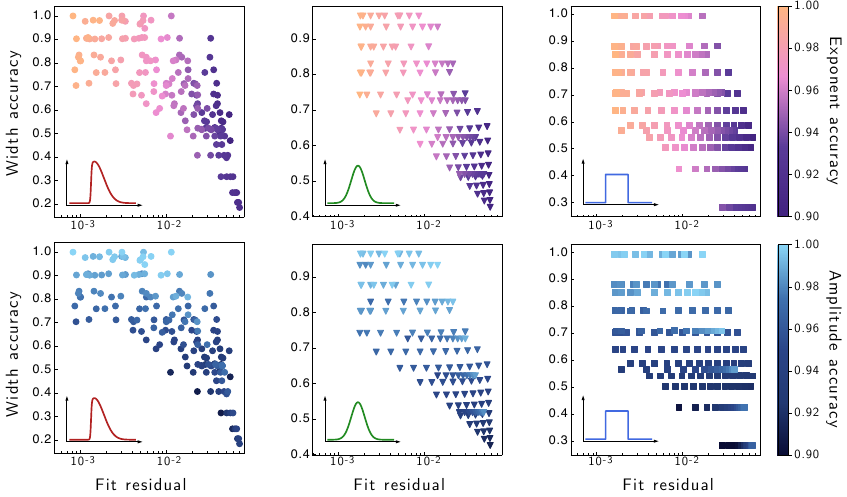}
    \caption{Scatter comparison between minimised fit residual and the width accuracy for the complete set of guess depth distributions. Columns correspond to different guess profile forms: left, Gauss-cutoff (ground truth); middle, Gaussian; right, rectangle. Colours represent fitted accuracy to the parameter values defining the truth power-law source spectrum: top, amplitude; bottom, amplitude.}
    \label{fig:fullscatter}
\end{figure}
\noindent In this work, we consider three types of depth profile:\\
\indent 1. The rectangular distribution
\begin{equation}
    P_\text{rect}(d; d_c, w)= 
    \begin{cases}
    \frac{1}{w} \qquad  &,\;\;d_c\leq d \leq d_c+w \\
    0\quad &,\;\;\text{otherwise}
    \end{cases}
    \label{eq:rect_dist}
\end{equation}
where $d_c$ is the short-depth cutoff and $w$ is the width of the distribution.\\
\indent 2. The Gaussian
\begin{equation}
    P_\text{gauss}(d; \mu, \sigma)=\exp\biggl[\frac{-(d-\mu)^2}{2\sigma^2}\biggr],
\end{equation}
where $\mu$ is the mean depth and $\sigma$ is the standard deviation.\\
\indent 3. The Gaussian-cutoff, a composite function defined as
\begin{align}
    P_\text{g-cut}(d;\mu,\sigma,d_c,s) &=P_\text{gauss}(d;\mu,\sigma)\cdot P_\text{sigmoid}(d;,d_c,s)\\
    &=\exp\biggl[\frac{-(d-\mu)^2}{2\sigma^2}\biggr]\cdot \frac{1}{1+\exp[-s(d-d_c)]},
\end{align}
where $\mu,\sigma$ are defined as previous, $d_c$ is the short-depth cutoff to the sigmoid, and $s$ defines the sharpness of the slope. For the purpose of this study, we elect to fix $s=8e9$ to give a smoothed but sharp ($<$1~nm) cutoff over the nm-scale depth profiles considered.\\\\
The truth dataset fitted in Fig.~\ref{fig:fullscatter} is generated using a Gauss-cutoff sensor distribution defined by the parameters $\{\mu =8~\text{nm},\sigma =3~\text{nm}, d_c = 7~\text{nm},s =8e9\}$, and a power-law source spectrum ($10^C\omega^{-\alpha}$) defined as $\{\alpha =1,C =11\}$ at a nominal depth $d_\text{nom}=8$~nm. Guess cases are generated using combinations on coarse sampling around each parameter defining the guess profile. The exact sampled sets are shown in tables below:
\begin{table}[h]
\centering
\caption{Sampled Gauss-cutoff parameters}
\begin{tabular}{|c|c|c|c|}
\hline
\textbf{Parameters} & \textbf{Min [nm]} & \textbf{Max [nm]} & \textbf{$N_\text{samples}$} \\
\hline
$\mu$ & 6 & 11 & 6 \\
\hline
$\sigma$ & 1 & 5 & 5 \\
\hline
$d_c$ & 4 & 9 & 6 \\
\hline
\end{tabular}
\end{table}

\begin{table}[h]
\centering
\caption{Sampled Gaussian parameters}
\begin{tabular}{|c|c|c|c|}
\hline
\textbf{Parameters} & \textbf{Min [nm]} & \textbf{Max [nm]} & \textbf{$N_\text{samples}$} \\
\hline
$\mu$ & 8 & 12 & 9 \\
\hline
$\sigma$ & 1 & 5 & 21 \\
\hline
\end{tabular}
\end{table}

\begin{table}[h]
\centering
\caption{Sampled rectangle parameters}
\begin{tabular}{|c|c|c|c|}
\hline
\textbf{Parameters} & \textbf{Min [nm]} & \textbf{Max [nm]} & \textbf{$N_\text{samples}$} \\
\hline
$d_c$ & 4 & 10 & 13 \\
\hline
$w$ & 2 & 14 & 13 \\
\hline
\end{tabular}
\end{table}

\subsection{Distinction factor definition}
\begin{figure}[H]
    \centering
    \includegraphics[width=\linewidth]{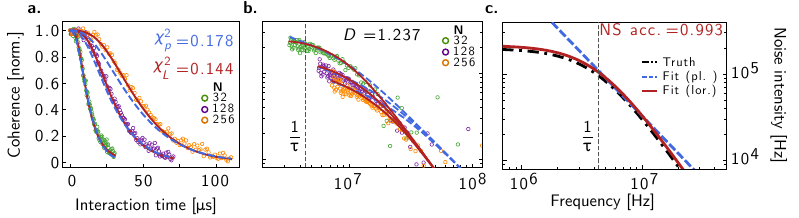}
    \caption{Distinction factor definition \textbf{a.} $\chi^2$ values are calculated from residuals between input coherence data (here for 2\% noise) and corresponding guess curves generated by the best fits to $S$ when guessing a Lorentzian ($\chi^2_L$) or power-law ($\chi^2_P)$ form. \textbf{b.} Comparison of subsequent $\bar{S}$, where improved residuals for a (correct) Lorentzian guess is quantified by a distinction factor, $D=\chi^2_P/\chi^2_L$ \textbf{c.} Fitted source spectra $S'$ compared to truth.}
    \label{fig:distinctionfactor}
\end{figure}
To quantify the ability to discern to true shape of the underlying source spectrum $S'$, we defined a distinction factor in the main text (Fig.~4c, main text) based on data residuals to a pseudo-experimental dataset. The quantity, $D$, is the ratio
\begin{equation}
    D=\frac{\chi^2_\text{wrong}}{\chi^2_\text{correct}},
\end{equation}
where $\chi^2_\text{correct}$ ($\chi^2_\text{wrong}$) is the chi-squared residual between the experimental coherence data and the corresponding coherence curves generated from the best fit parameters of a correct (wrong) guess spectral shape. An example of this procedure is displayed in Fig.~\ref{fig:distinctionfactor}, where $\chi^2$ values are calculated from the best fits of $S$ when the depth distribution is known (Fig.~\ref{fig:distinctionfactor}a). The fits to $\bar{S}$ is shown explicitly in Fig.~\ref{fig:distinctionfactor}b, where the accuracy of the two fits corroborate with the accuracy of the guess form. Fig.~\ref{fig:distinctionfactor}c compares the best fits to the parameterised source spectrum $S'(\omega)$, compared to the truth. The correct (Lorentzian) guess closely aligns with the truth (a total accuracy above 99\%), while the wrong (power-law) guess deviates with tangential agreement with the truth around its inflection point, which falls within the measured data range ($\sim$40-100~MHz).

\subsection{Correspondence between fitted accuracy and distinction factor}
\begin{figure}[H]
    \centering
    \includegraphics[width=0.9\linewidth]{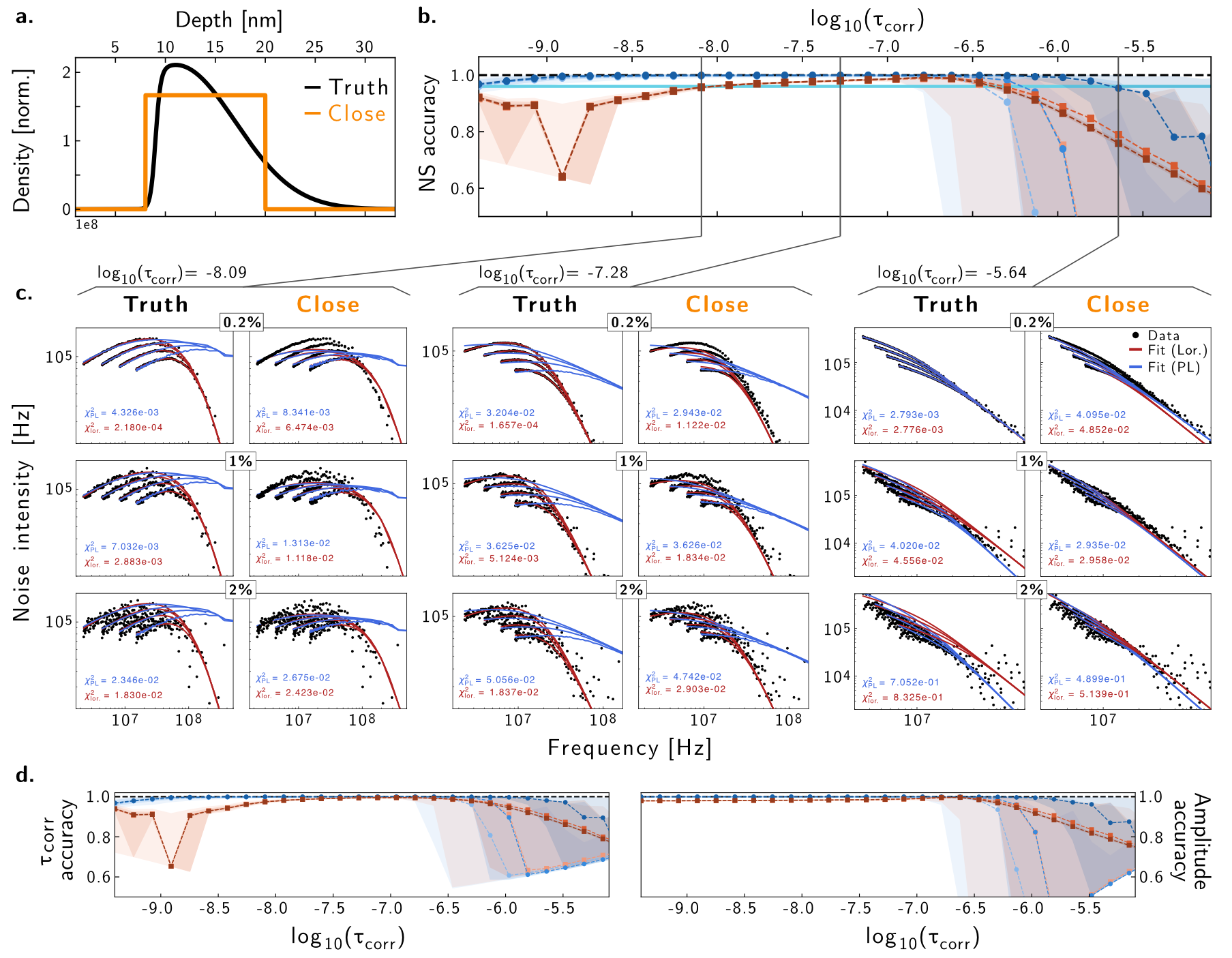}
    \caption{\textbf{a.} Truth depth distribution (black) used to generate simulated experimental data compared to a \textit{close} guess for the distribution (yellow) used for fitting. \textbf{b.} Accuracy of source noise spectrum fit as the $1/\tau_\text{corr}$ inflection point of the true Lorentzian is swept across the measured range. Error bars represent 2$\sigma$ variation from $N_\text{runs}=100$ Monte-Carlo simulations of fitted values (reproduced from Fig.~4c). \textbf{c.} Simulated experimental data compared to power-law (blue) and Lorentzian (red) fits for varying noise levels on explicit $\tau_\text{corr}$ cases. Cases are selected where the inflection point is (middle) best measured, (right) not seen at all, or sitting at an intermediate threshold (left). \textbf{d.} Accuracies of individual parameters describing the Lorentzian: left, $\tau_\text{corr}$; right, amplitude.}
    \label{fig:3fexplicit}
\end{figure}
An ensemble noise spectrum dataset was generated with a gauss-cutoff depth distribution coupling a Lorentzian noise spectrum with a varying characteristic $\tau_\text{corr}$ value. The data is fit with both knowledge of the underlying truth and an incorrect but `close' rectangular guess (as characterised by $>$96\% total accuracy in Fig.~3c main text); the distributions of which are shown in Fig.~\ref{fig:3fexplicit}a. The total accuracy of the Lorentzian source spectrum reconstruction for both guesses is shown in Fig.~\ref{fig:3fexplicit}b (reproduced from Fig.~4c main text) for three levels of noise. Fitting with precise knowledge of the true depth profile yields high fitting accuracy when the inflection point of the source spectrum falls within the measured frequency range. With a close guess, although maximal accuracy is prevented by an inexact profile, the trend of sustained high accuracy over central $1/\tau_\text{corr}$ values is much the same. The separated accuracy plots for each fitted parameter present expected behaviour. For inflection points in the high-$f$ (low $\tau_\text{corr}$) limit, the $\tau_\text{corr}$ value is difficult the fit while the amplitude of the Lorentzian, which takes on a flat ($\alpha=0$) power-law-like spectrum over the measured range, is accurately fit. In the low-$f$ (high $\tau_\text{corr}$) limit however, both parameters are inaccurately fit as the probed range is blind to both central amplitude and inflection point when measuring the $\alpha=2$ power-law tail of the Lorentzian.

\section{Application to experiment}

\subsection{Depth distribution fitting via NV-NMR}
\label{sec:NMR_DDfitting}
\begin{figure}[H]
    \centering
    \includegraphics[width=\linewidth]{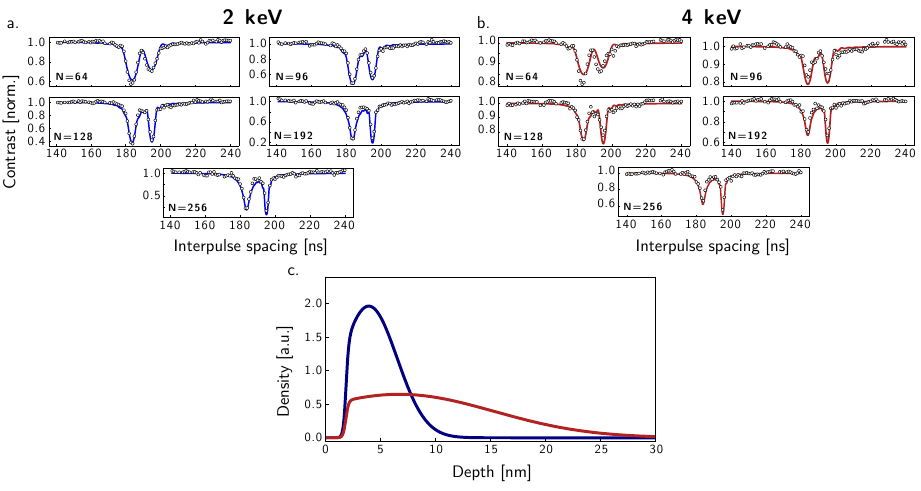}
    \caption{Series of XY8-based NMR datasets used to fit a depth distribution for the {a.} 2~keV and \textbf{b.} 4~keV ensembles. Background-free experimental data are shown in black. Contrast curves predicted from the globally fit depth profiles (shown explicitly in \textbf{c.}) are represented by the coloured traces (blue, 2~keV; red, 4~keV).}
    \label{fig:NMRfitting}
\end{figure}
\noindent Pham \textit{et al.} \cite{pham2016} provide a mathematical description for the NMR signal measured from a single sensor coupled to a target spin geometry at its host interface. The measurable contrast $C(\tau)$, driven by phase accumulation from a gaussian spin noise environment, can be written as
\begin{equation}
    C(\tau)\approx\exp\biggl\{-\frac{2}{\pi^2}\gamma_e^2B_{RMS}^2(d_{NV})K(\tau, N)\biggr\},
    \label{eq:NMRcontrast}
\end{equation}
where $\gamma_e$ is the NV gyromagnetic ratio, $B_{RMS}$ relates to the magnetic signal originating from nuclear spins, and $K$ encodes the filtering of the dynamical decoupling sequence modulating the NV spin.

The filter function $K$ is computed from the parameters defining the sequence: pulse number N and interpulse spacing $\tau$. In the consideration of a finite nuclear spin dephasing time, $T_{2,\text{nuc}}^*$, $K(\tau,N)$ evaluates as
\begin{align}
K(N\tau) &\approx \frac{2T_{2n}^{*2}}{\left[1 + T_{2n}^{*2}\left(\omega_L - \dfrac{\pi}{\tau}\right)^2\right]^2} \notag\\
&\quad \times \Bigg( e^{-\frac{N\tau}{T_{2n}^{*}}} \Bigg\{ \left[1 - T_{2n}^{*2}\left(\omega_L - \frac{\pi}{\tau}\right)^2\right] \cos\left[N\tau\left(\omega_L - \frac{\pi}{\tau}\right)\right] \notag\\
&\qquad - 2T_{2n}^{*}\left(\omega_L - \frac{\pi}{\tau}\right)\sin\left[N\tau\left(\omega_L - \frac{\pi}{\tau}\right)\right] \Bigg\} \notag\\
&\qquad + \frac{N\tau}{T_{2n}^{*}}\left[1 + T_{2n}^{*2}\left(\omega_L - \frac{\pi}{\tau}\right)^2\right] + T_{2n}^{*2}\left(\omega_L - \frac{\pi}{\tau}\right)^2 - 1 \Bigg),
\end{align}
where $\omega_L$ is the Larmor frequency of the target nuclear species. The $B_\text{RMS}$ is calculated from the system geometry, of which there are two relevant to this work: a thin layer (i.e. \ce{^1H} adsorbates) and a semi-infinite volume (i.e. \ce{^19F} Fomblin oil liquid coating). For an NV center oriented at an angle $\alpha = 54.7^\circ$, as is the case for the [100]-diamond substrate used in this work, the $B_\text{RMS}$ expressions for these cases are given by (see Pham \textit{et al.} \cite{pham2016}) for derivations):
\begin{align}
    B_\text{RMS,semi-inf}^2&=\rho\biggl(\frac{\mu_0\hbar\gamma_n}{4\pi}\biggr)^2\frac{5\pi}{96d_{NV}^3}\\
    B_\text{RMS,thin}^2&=\frac{5\pi\rho}{96}\biggl(\frac{\mu_0\hbar\gamma_n}{4\pi}\biggr)^2\biggl(\frac{1}{d_{NV}^3}-\frac{1}{(d_{NV}+t)^3}\biggr)
\end{align}
Here $\rho$ is the nuclear density of the sample, $\gamma_n$ is the target spin gyromagnetic ratio, and $d_{NV}$ is the depth of the NV from the interface of the spin layer. In the case of the thin layer, $t$ is the layer thickness. With precise knowledge of the target species and its density, it is therefore possible to invert the measured NMR contrast reported by a shallow sensor for its depth.

Using Eq.~\ref{eq:NMRcontrast} with the appropriate spin geometry for $B^2_\text{RMS}$ and knowledge of the layer densities and relevant thicknesses, the expected background-free NMR signal can be simulated for an NV at some depth $d$ beneath the surface. This can be extended to an ensemble signal by assuming that the ensemble averaged contrast $\bar{C}$ comprises weighting contributions based on the probability density function $P(d)$ associated with the ensemble depth distribution
\begin{equation}
    \bar{C} = \int_0^\infty P(d')C(d')dd'
\end{equation}
Owed to the idea that spin contrast is accumulated differently across an ensemble depending on the pulse sequence number $N$ (see \cite{ziem2019} or Fig.~2a, main text), a series of $\bar{C}_N$ therefore forms a basis to which a depth distribution may be fit.

The distribution is fit via a least-squares minimisation procedure, where the measured basis of background-normalised experimental \{$\bar{C}_N$\} curves is used to fit a global depth distribution $P(d)$ for its parameters. The objective function residual is defined as $C_\text{exp}/C_\text{guess}$ to balance contributions across dataset elements which vary in dip contrast. Using this method, the depth distribution parameters as well as the thickness(es) of the intermediate spin layer(s) can be fit.

The best fits using this method are shown in Fig.~\ref{fig:NMRfitting}. Here the NV$^-$ ionisation cut-off value $d_c$ is determined from the 2~keV ensemble and subsequently held as a fixed parameter in a constrained fit of the 4~keV ensemble distribution. We reduce the parameter space in this way to aid converge and resolve fit degeneracies in the case of a deeper ensemble with lower surface sensitivity.

parameter space of the fitting to coach convergence, and disambiguate the fit in the case of a less surface-sensitive ensemble.

\subsection{Decoherence curve fitting}
\begin{figure}[H]
    \centering
    \includegraphics[width=0.9\linewidth]{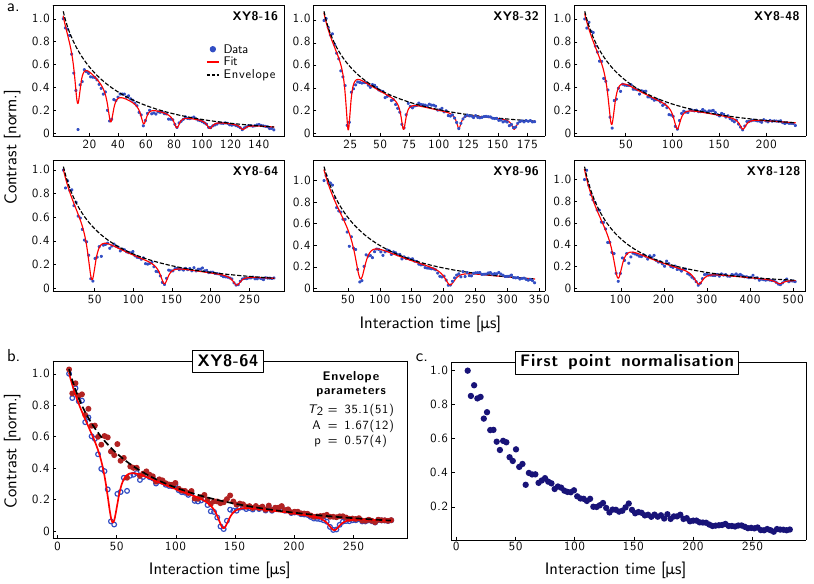}
    \caption{\textbf{a.} Composite $T_2$ fits for all XY8-$N$ decoherence curves comprising the 4~keV dataset. Reference-normalised data (blue, dots) is fit for both the stretched exponential decay envelope (black, dashed) and the relevant Lorentzian \ce{^13C} harmonic revivals (composite fit shown in red) present across the measured interaction time range. An example ($N$ = 64) of the process to first \textbf{b.} subtract \ce{^13C} revivals and then \textbf{c.} first-point normalise using the fitted composite function}
    \label{fig:C13fitting}
\end{figure}
Generally, a series of XY8-$N$ measurements are taken with increasing pulse number (here $N=[16,32,48,64,96,128]$) to increase the number of data points extend the sweep range of the filter function selection window. For ensemble noise spectroscopy, the emergent striations help to disentangle the influence of the depth profile from the source spectrum. 

Reference-normalised experimental data taken from the 4~keV ensemble is shown in Fig.~\ref{fig:C13fitting}a. Due to the natural abundance (1.1\%) of isotopic \ce{^13C} in the host diamond substrate, experimental decoherence curves will exhibit sharp dips (or \ce{^13C} revivals) in coherence at $\tau$ points corresponding to integer multiples of the \ce{^13C} Larmor frequency. We deal with these revivals by fitted a composite function comprising a series of Lorentzians enveloped by a stretched exponential. The resonance of each Lorentzian is constrained by its prediction via the external magnetic field strength ascertained from optically-detected magnetic resonance (ODMR):
\begin{equation}
    f_{^{13}C}=(2n+1) \times10.7084\;[\text{MHz}\;\text{T}^{-1}]\times B_\text{ext}\;[\text{T}],\;\quad n\in\mathbb{N}
\end{equation}
All predicted peaks along the decoherence curves ($T=N/2f$) are fit for their amplitude as well as a global width, providing a fit curve that fits experimental data well. The reference-normalised data is additively rescaled by the difference between the envelope and the composite function at every point to remove contributions from \ce{^13C} revivals. An example of this is shown in Fig.~\ref{fig:C13fitting}b. The data is lastly divided by the envelope to ensure consistent first-point normalisation in the presence of experiment noise (Fig.~\ref{fig:C13fitting}c).
\subsection{Experimental noise spectrum fits for different spectral shapes}
\label{sec:allNSfits}
\begin{figure}[H]
    \centering
    \includegraphics[width=\linewidth]{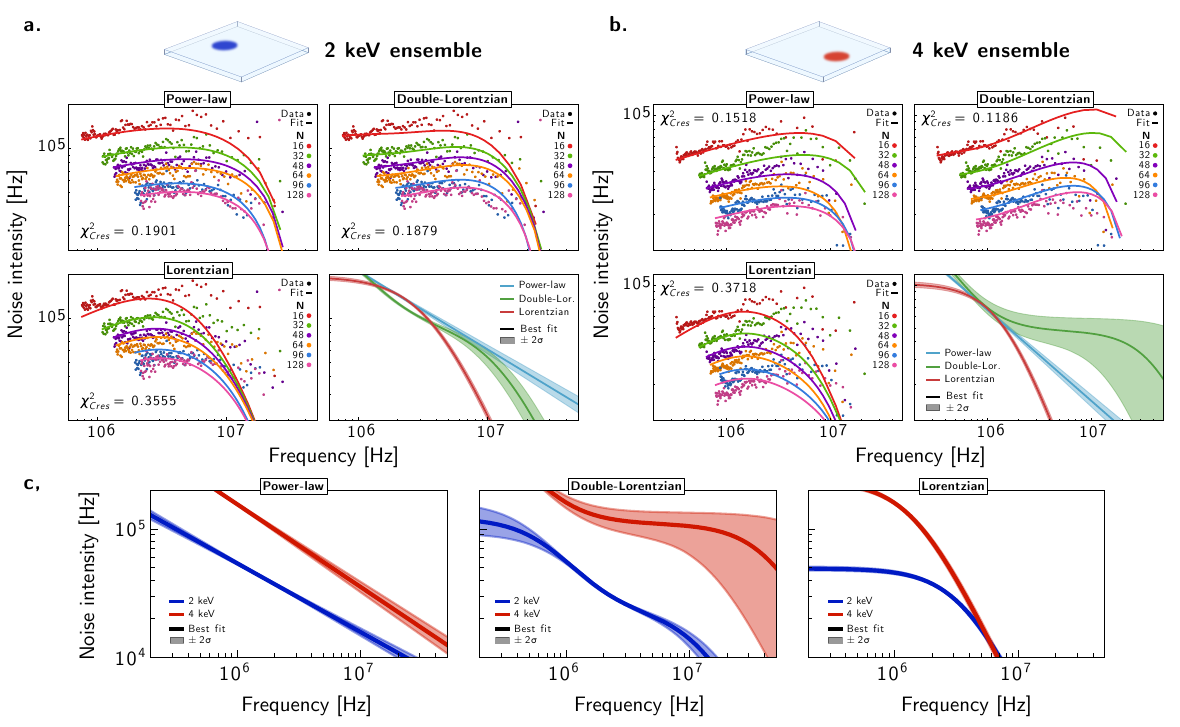}
    \caption{Best fits for the three noise spectral shapes considered in this work, for data taken from the \textbf{a.} 2~keV ensemble and \textbf{b.} 4~keV ensemble. Fits to data are shown for each of the three cases, while the bottom right panel compares the fitted source spectrum for all $S'$ forms considered, evaluated at the corresponding mean depth of the ensemble. \textbf{c.} Comparison of the surface noise spectrum $S'$ fitted by each ensemble, evaluated at the same nominal depth $d_\text{nom} =8$~nm. Each panel shows a different guess spectral shape.}
    \label{fig:fig4_fulldata}
\end{figure}

In Fig.~\ref{fig:fig4_fulldata}, we show the full fit comparisons for each of the three source spectrum shapes considered in this work. Here, fits to the experimental noise spectra $\bar{S}$ as measured by the 2~keV (Fig.~\ref{fig:fig4_fulldata}a) and 4~keV (Fig.~\ref{fig:fig4_fulldata}b) ensembles are shown for a power-law (top left), double-Lorentzian (top right), and Lorentzian (bottom left) guess to the source spectrum $S'$. From these plots, we can see that both a power-law and double-Lorentzian guess for the shape of $S'$ leads to a good fit for the data while the best-fit case for the Lorentzian does not. These observations are reflected in the fit residuals, where the $\chi^2$-value for a Lorentzian guess is close to double that of the other two guesses, which are relatively similar. All source spectrum fits are also shown, evaluated at the ensemble's mean depth (bottom right). Here we can see that the similar residual between the power-law and double-Lorentzian, which suggestive of the fit accuracy when the depth distribution is well constrained (Fig.~4, main text), is corroborated with a similarity between $S'$ spectra over the measurable range.

In Fig.~\ref{fig:fig4_fulldata}c we plot the reconstructed source spectra for the two ensembles evaluated instead at the same nominal depth, $d_\text{nom} = 8$~nm, to directly compare the resolved intensity of the surface noise spectrum between ensembles. We find that for all considered source spectral shapes, the intensity of noise coupling to the 4~keV ensemble is larger than that experienced by the 2~keV ensemble. A possible explanation for this is a bulk contribution we do not account for of similar intensity to the surface noise (see Sec.~\ref{sec:bulkcontribution}). This contribution would have highly prevalence in the 4~keV sample because it is a deeper profile. Additionally, this bulk contribution may not be uniform across the ensemble -- it being related to expected populations of unconverted nitrogen and vacancy clusters which are correlated to the ion straggle statistics. In this case, it may appear not a fixed offset but as a secondary component with its own distribution of coupling that could get erroneously fit by our procedure, possibly giving rise to the high frequency cutoff seen in the double-Lorentzian fit to the 4~keV dataset. In principle it is possible to account for such effects in our procedure at the cost of added complexity and extra free parameters, however we leave this for future work.

\subsection{Coupling strength scaling with depth}
\label{sec:dscaling}

\begin{figure}[H]
    \centering
    \includegraphics[width=\linewidth]{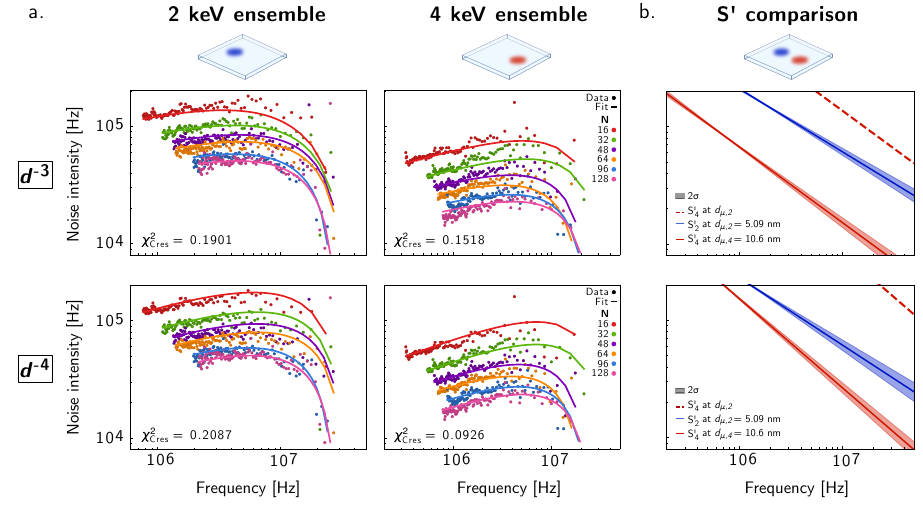}
    \caption{\textbf{a.} Ensemble noise spectrum fits for the 2~keV (left) and 4~keV (right) experimental data, using a (top) $d^{-3}$ or (bottom) $d^{-4}$ noise intensity scaling with depth. \textbf{b.} Reconstructed source spectrum $S'$ fits from 2~keV ($S'_2$, blue) and 4~keV ($S'_4$, red) data fitting, evaluated at their respective mean depth $d_{\mu}$. The corresponding amplitude of $S'_4$ evaluated at $d_{\mu,2}$ is also shown (dashed).}
    \label{fig:dcsaling}
\end{figure}

In the main text, we fit the surface noise features with a $d^{-3}$ scaling, which is applicable to detecting magnetic noise signal from a semi-infinite volume at the diamond surface. If noise is truly two-dimensional however, a $d^{-4}$ scaling would instead be expected. We consider this scaling in Fig.~\ref{fig:dcsaling}. We find that the experimental data is fit well in both cases. Given surface roughness of the diamond and the possibility of adsorbates delocalising magnetic noise from the diamond surface, it is reasonable that the origins of surface noise are not truly two-dimensional compared to the extent of the NV sensing layers. We chose to present the $d^{-3}$ scaling in the main text for consistency with the rest of the text and since the extracted power law exponents for the 2 and 4~keV data sets align more closely: $\alpha_2=0.54(2)$ and $\alpha_4=0.65(2)$ respectively for $d^{-3}$ scaling, while $\alpha_2=0.57(3)$ and $\alpha_4=0.76(2)$ for $d^{-4}$ scaling. Note that the 2~keV fits are similar (within error) in both cases.

\bibliography{library}